\documentclass[twocolumn,tighten,times, trackchanges]{aastex701}

\usepackage{mathrsfs}
\newcommand{\feii}{\ifmmode {\rm Fe\ II} \else Fe~{\sc ii}\fi}
\newcommand{\heii}{\ifmmode {\rm He\ II} \else He~{\sc ii}\fi}
\newcommand{\hei}{\ifmmode {\rm He\ I} \else He~{\sc i}\fi}
\newcommand{\oiii}{\ifmmode {\rm~[O\ III]} \else [O~{\sc iii}]\fi}
\newcommand{\nii}{\ifmmode {\rm~[N\ II]} \else [N~{\sc ii}]\fi}
\newcommand{\mgii}{\ifmmode {\rm~Mg\ II} \else Mg~{\sc ii}\fi}
\newcommand{\ciii}{\ifmmode {\rm~C\ III]} \else C~{\sc iii}]\fi}
\newcommand{\civ}{\ifmmode {\rm~C\ IV} \else C~{\sc iv}\fi}
\newcommand{\hb}{\rm{H$\beta$}}
\newcommand{\ha}{\rm{H$\alpha$}}
\newcommand{\hg}{\rm{H$\gamma$}}
\newcommand{\hbred}{\rm{H$\beta_{\rm red}$}}
\newcommand{\hbcor}{\rm{H$\beta_{\rm core}$}}
\newcommand{\mbh}{\ifmmode {M_{\bullet}} \else $M_{\bullet}$\fi}
\def\mathdotM{\dot{\mathscr{M}}}
\def\sunm{M_{\odot}}

\def\ergs{\rm erg\,s^{-1}}
\def\ergscm{\rm erg\,s^{-1}\,cm^{-2}}
\def\ergscma{\rm erg\,s^{-1}\,cm^{-2}\,\AA^{-1}}
\newcommand{\kms}{$\rm km\,s^{-1}$}
\newcommand{\Rfe}{\ifmmode {{\cal R}_{\rm Fe}} \else ${\cal R}_{\rm Fe}$\fi}

\begin{document}

\title{The Most Luminous H$\beta$ Reverberation Mapping of E1821+643 Indicates the Lower Boundary of the Radius-Luminosity Relation}
\shorttitle{Reverberation Mapping of E1821+643}

\shortauthors{Li et al.}

\author[0000-0003-3823-3419]{Sha-Sha Li}
\affiliation{Yunnan Observatories, Chinese Academy of Sciences, Kunming 650216, Yunnan, People's Republic of China}
\affiliation{Key Laboratory for the Structure and Evolution of Celestial Objects, Chinese Academy of Sciences, Kunming 650216, Yunnan, People's Republic of China}
\affiliation{Center for Astronomical Mega-Science, Chinese Academy of Sciences, 20A Datun Road, Chaoyang District, Beijing 100012, People's Republic of China}
\affiliation{Key Laboratory of Radio Astronomy and Technology, Chinese Academy of Sciences, 20A Datun Road, Chaoyang District, Beijing 100101, People's Republic of China}
\email[]{lishasha@ynao.ac.cn} 

\author[0000-0002-1530-2680]{Hai-Cheng Feng}
\affiliation{Yunnan Observatories, Chinese Academy of Sciences, Kunming 650216, Yunnan, People's Republic of China}
\affiliation{Key Laboratory for the Structure and Evolution of Celestial Objects, Chinese Academy of Sciences, Kunming 650216, Yunnan, People's Republic of China}
\affiliation{Center for Astronomical Mega-Science, Chinese Academy of Sciences, 20A Datun Road, Chaoyang District, Beijing 100012, People's Republic of China}
\affiliation{Key Laboratory of Radio Astronomy and Technology, Chinese Academy of Sciences, 20A Datun Road, Chaoyang District, Beijing 100101, People's Republic of China}
\email[show]{hcfeng@ynao.ac.cn}

\author[0000-0002-2581-8154]{Jiancheng Wu}
\affiliation{Institute for Astronomy, School of Physics, Zhejiang University, 866 Yuhangtang Road, Hangzhou 310058, People’s Republic of China}
\affiliation{Department of Astronomy, School of Physics, Huazhong University of Science and Technology, Luoyu Road 1037, Wuhan, China}
\email[]{}

\author{J. M. Bai}
\affiliation{Yunnan Observatories, Chinese Academy of Sciences, Kunming 650216, Yunnan, People's Republic of China}
\affiliation{Key Laboratory for the Structure and Evolution of Celestial Objects, Chinese Academy of Sciences, Kunming 650216, Yunnan, People's Republic of China}
\affiliation{Center for Astronomical Mega-Science, Chinese Academy of Sciences, 20A Datun Road, Chaoyang District, Beijing 100012, People's Republic of China}
\affiliation{Key Laboratory of Radio Astronomy and Technology, Chinese Academy of Sciences, 20A Datun Road, Chaoyang District, Beijing 100101, People's Republic of China}
\email[]{}

\author[0000-0002-2153-3688]{H. T. Liu}
\affiliation{Yunnan Observatories, Chinese Academy of Sciences, Kunming 650216, Yunnan, People's Republic of China}
\affiliation{Key Laboratory for the Structure and Evolution of Celestial Objects, Chinese Academy of Sciences, Kunming 650216, Yunnan, People's Republic of China}
\affiliation{Center for Astronomical Mega-Science, Chinese Academy of Sciences, 20A Datun Road, Chaoyang District, Beijing 100012, People's Republic of China}
\email[]{}

\author[0000-0002-2310-0982]{Kai-Xing Lu}
\affiliation{Yunnan Observatories, Chinese Academy of Sciences, Kunming 650216, Yunnan, People's Republic of China}
\affiliation{Key Laboratory for the Structure and Evolution of Celestial Objects, Chinese Academy of Sciences, Kunming 650216, Yunnan, People's Republic of China}
\affiliation{Center for Astronomical Mega-Science, Chinese Academy of Sciences, 20A Datun Road, Chaoyang District, Beijing 100012, People's Republic of China}
\email[]{}

\author[0000-0002-0771-2153]{Mouyuan Sun}
\affiliation{Department of Astronomy, Xiamen University, Xiamen, Fujian 361005, People’s Republic of China}
\email[]{}

\author[0000-0003-4156-3793]{Jian-Guo Wang}
\affiliation{Yunnan Observatories, Chinese Academy of Sciences, Kunming 650216, Yunnan, People's Republic of China}
\affiliation{Key Laboratory for the Structure and Evolution of Celestial Objects, Chinese Academy of Sciences, Kunming 650216, Yunnan, People's Republic of China}
\affiliation{Center for Astronomical Mega-Science, Chinese Academy of Sciences, 20A Datun Road, Chaoyang District, Beijing 100012, People's Republic of China}
\email[]{}

\begin{abstract}
The radius-luminosity ($R_{\rm BLR}$-$L_{5100}$) relation is fundamental to active galactic nucleus (AGN) studies, enabling supermassive black hole (SMBH) mass estimates and AGN-based cosmology applications. However, its high-luminosity end remains poorly calibrated due to insufficient reliable reverberation mapping (RM) data. We present a four-year RM campaign of the luminous quasar E1821+643 using the Lijiang 2.4-m telescope, supplemented by archival multi-wavelength data. E1821+643 is the most luminous AGN with an \hb\ RM measurement to date. The measured time lag of $83.2_{-18.7}^{+17.5}$ days is a factor of 5.6 shorter than predicted by the canonical $R_{\rm BLR}$-$L_{5100}$ relation. By compiling the full \hb\ RM sample, we find that such deviation defines a lower envelope ($0.2R_{\rm BLR}$) of measured lags across the entire luminosity range, while the upper envelope lies near $2R_{\rm BLR}$, implying that the scatter for individual AGNs can reach 1 dex. Spectral decomposition reveals two distinct \hb\ components: a core component with a lag of $267.0_{-17.6}^{+16.6}$ days closer to the $R_{\rm BLR}$-$L_{5100}$ relation, and a redshifted tail with a much shorter lag of $-49.0_{-34.5}^{+50.5}$ days. The short-lag component not only accounts for the significantly shortened overall lag, but also leads to an opposite interpretation of the intrinsic BLR kinematics. These effects can introduce systematic uncertainties in black hole mass estimates by factors of up to tens. Our findings demonstrate that shortened lags in high-accretion-rate AGNs arise from multi-component BLR structures, posing substantial challenges to single-epoch mass estimates and impacting SMBH demographics and cosmological applications.
\end{abstract}


\section{Introduction} \label{sec:1}
The broad-line region (BLR) of active galactic nuclei (AGNs) consists of dense, ionized gas clouds located in the vicinity of the central supermassive black hole (SMBH). These clouds move at velocities ranging from several thousand to tens of thousands of \kms\ under the SMBH’s gravitational potential, producing the characteristic broad emission lines observed in AGN spectra via Doppler broadening. As the only region within the inner structure of AGNs where both the geometry and kinematics can be measured, the BLR provides a unique probe for measuring the mass of the central SMBH.

Direct spatial resolution of the BLR remains difficult in most cases due to its compact size, except for a handful of AGNs successfully resolved by the GRAVITY interferometer \citep{GRAVITY2024}. Alternatively, the most widely used technique for probing the BLR and estimating SMBH masses is reverberation mapping (RM), which measures the time delay between variations in the AGN continuum and the broad-line response \citep{Blandford1982, Peterson1993}. This time lag ($\tau$) corresponds to the light-travel time across the BLR and provides a characteristic radius: $R_{\rm BLR} = c\tau$. Combined with the velocity ($v$) of the BLR gas, inferred from the emission-line width, this yields a black hole mass estimate: 
\begin{equation}\label{equ1}
\mbh = f \frac{c\tau v^2}{G}, 
\end{equation}
where $f$ is a dimensionless virial factor that accounts for the geometry and kinematics of the BLR \citep{Li2025}. Over the past four decades, RM has provided SMBH mass estimates for more than 150 AGNs \citep[e.g.,][]{Bentz2010, Feng2021a, Feng2024, Villafana2022, Cho2023, Woo2024}.

A key discovery from RM studies is the empirical correlation between the BLR size and AGN luminosity, known as the radius-luminosity ($R_{\rm BLR}-L_{\rm 5100}$) relation \citep{Kaspi2000}, where $R_{\rm BLR} \propto L_{\rm 5100}^{\alpha}$ with $\alpha \approx 0.5$ \citep{Bentz2013}. This relation arises from the photoionization physics of the BLR, in which the size of the line-emitting region scales with the ionizing flux from the central engine. The $R_{\rm BLR}-L_{\rm 5100}$ relation has significant implications for black hole mass measurements, particularly for AGNs at high redshifts \citep{Wu2015, Yang2021} where RM is impractical. It enables the estimation of $R_{\rm BLR}$ from single-epoch spectra, bypassing the need for time-intensive monitoring. Combined with the broad-line width to infer the virial velocity, this method allows single-epoch black hole mass estimates for large AGN samples and has become a cornerstone of AGN studies and cosmological applications \citep[e.g.,][]{Negrete2018, Panda2024, Cao2025}.

Despite its fundamental importance, the current $R_{\rm BLR}-L_{\rm 5100}$ relation suffers from two issues that limit its accuracy and applicability. First, the intrinsic scatter of the relation has not been consistently quantified. \citet{Bentz2013} reported a remarkably tight correlation with a scatter of 0.13 dex. However, subsequent RM campaigns reveal that high-accretion-rate AGNs systematically exhibit shorter lags than predicted \citep{Du2016, Du2018, Yao2024}, though the physical origin of this deviation remains unclear. When these objects are included, the observed scatter can increase to 0.2-0.3 dex \citep{Du2019, Wang2024}, and in some individual cases, the lags deviate from the expected values by factors of several \citep{Li2021, Lu2022, Feng2025a}. Second, the high-luminosity end of the relation is poorly constrained due to observational challenges. Luminous AGNs require long-term monitoring to measure their larger BLRs, making such observations resource-intensive. Consequently, only a few quasars have reliable lag measurements at $L_{\rm 5100} > 10^{45} \ergs$, introducing substantial uncertainties when extrapolating the $R_{\rm BLR}-L_{\rm 5100}$ relation to the luminous regime. This is particularly concerning because the majority of known AGNs populate this poorly calibrated range \citep[e.g.,][]{Rakshit2020}.

To address these critical uncertainties, we conduct an intensive four-year RM observation of E1821+643, a luminous quasar with a high accretion rate. Meanwhile, a sparse spectroscopic monitoring campaign has revealed an \hb\ lag shorter than predicted by the $R_{\rm BLR}$-$L_{\rm 5100}$ relation \citep{Shapovalova2016}. 

This paper is organized as follows. In Section~\ref{sec:2}, we describe the observations and data reduction. Section \ref{sec:3} presents our analysis and main results. In Section \ref{sec:4}, we discuss the implications of our findings. Finally, Section \ref{sec:5} summarizes our conclusions. Throughout this paper, we adopt the cosmology with $H_0 = 67$ km s$^{-1}$ Mpc$^{-1}$, $\Omega_m = 0.32$, and $\Omega_\Lambda=0.68$ \citep{PlanckCollaboration2020}.

\section{Observations and Data Reduction} \label{sec:2}
The quasar E1821+643 is monitored from March 2022 to October 2025 using the 2.4-m telescope at Lijiang Observatory, operated by the Yunnan Observatories of the Chinese Academy of Sciences. Observations are carried out with the Yunnan Faint Object Spectrograph and Camera (YFOSC), which enables both photometric and spectroscopic observations \citep{Wang2019, Xin2020}. Below, we briefly summarize the observations and the corresponding data reduction procedures.

\subsection{Photometry}
As part of our monitoring campaign, we obtained broadband imaging of E1821+643 using 50s exposures through a Johnson $B$ filter. The $B$ band was selected to minimize contamination from the host galaxy while maximizing the contrast of nuclear emission. The photometric data were reduced following standard procedures implemented in PyRAF \citep{Pyraf2012}, including bias subtraction and flat-field correction. We performed differential photometry using four stars within the field of view (FoV) to minimize systematic effects from atmospheric variations. A circular aperture radius of 3\farcs68 was adopted, with sky background estimated from an annular region with inner and outer radii of 11\farcs04 and 16\farcs70, respectively. These photometric measurements serve two main purposes: (1) to provide an independent verification of the spectroscopic flux calibration, and (2) to serve as continuum data for variability analysis.

\subsection{Spectroscopy}
For our spectroscopic observations, we initially employ Grism 3 with a 2\farcs5-wide slit. To mitigate secondary spectral contamination--arising from the overlap of second-order spectra at shorter wavelengths with first-order spectra at longer wavelengths--a UV-blocking filter is applied following the procedure outlined in \citet{Feng2020}. This configuration provided wavelength coverage from 4190 to 9000~\AA\ with a dispersion of 2.9~\AA\,pixel$^{-1}$, yielding five spectra during the initial phase.

Due to the redshift of E1821+643, the \ha\ emission line is redshifted beyond the optimal range of Grism 3. To achieve better coverage at longer wavelengths, we switch to Grism 8, which extends coverage to 5155-9500~\AA\ with improved dispersion of 1.5~\AA\,pixel$^{-1}$. Simultaneously, we widen the slit to 5\farcs05 to minimize flux losses under variable seeing conditions, while maintaining the UV-blocking filter. This optimized configuration became our primary setup, producing 114 spectra.

To enable direct comparison between the two grisms under identical conditions, we obtained three additional spectra using Grism 3 with the wider 5\farcs05 slit and UV-blocking filter. Their flux calibration prove consistent across the overlapping wavelength regions between the two configurations. We also acquired two additional spectra using Grism 3 with the 5\farcs05 slit but without the UV-blocking filter, extending wavelength coverage to 3500-9000~\AA\ for comprehensive stellar template construction. 

Other aspects of the observational strategy and data reduction procedures are consistent with those described in \citet{Feng2025a}. In short, the long-slit design enables simultaneous observation of the target and a nearby comparison star. Standard PyRAF routines are used for bias subtraction, flat-fielding, wavelength calibration via arc lamps, and spectral extraction using a 4\farcs24 aperture, with background regions defined between 14\farcs15 and 19\farcs81. Flux calibration and telluric correction are performed using the comparison star.

Our calibration strategy allows for reliable flux calibration even under poor weather conditions. However, occasional instrumental issues or sudden cloud cover during observations can lead to guiding failures, extremely low signal-to-noise (S/N) spectra, or distorted spectral profiles, thereby introducing outliers into the light curves. To exclude these problematic exposures, we adopt the approach described in \citet{Zhou2025}, applying a smoothing filter to the light curves and rejecting data points that deviate by more than 3$\sigma$. Note that this method may remove a few valid data points near temporal gaps. We therefore perform another independent quality check: (1) we examine the centroid positions of the \oiii\ emission line in the object spectra and absorption features in the comparison star spectra, and reject exposures with offsets exceeding 4 pixels ($\sim$1\farcs1, or 2 pixels for a 2\farcs5 slit); (2) we visually inspect each spectrum and exclude those with significantly distorted line profiles; and (3) we remove spectra with S/N $<$ 20 pixel$^{-1}$ near rest-frame 5100\AA. 

After applying these criteria, we exclude 14 low-quality exposures, all of which satisfy the rejection criteria defined by \citet{Zhou2025}. The final dataset consists of 110 spectra with an average S/N of 44 pixel$^{-1}$ near rest-frame 5100\AA.

\subsection{Archival Photometric Data}
To supplement our Lijiang observations, extend the temporal baseline and sampling cadence of the light curves, and probe the hot dust-emitting regions, we incorporate archival photometric data from a range of time-domain surveys spanning optical to mid-infrared (MIR) wavelengths. The optical data come from the Catalina Real-Time Transient Survey \citep[CRTS;][]{Drake2009}, the Palomar Transient Factory \citep[PTF;][]{Law2009}, the Zwicky Transient Facility \citep[ZTF;][]{Masci2019}, the All-Sky Automated Survey for Supernovae \citep[ASAS-SN;][]{Shappee2014}, and the Asteroid Terrestrial-impact Last Alert System \citep[ATLAS;][]{Tonry2018}, while the MIR data are obtained from the Wide-field Infrared Survey Explorer \citep[WISE;][]{Wright2010, Mainzer2011}.

CRTS provides unfiltered optical photometry calibrated to the Johnson $V$ band. PTF observes in the $g$ and $r$ bands, and its successor ZTF further extends this coverage to include the $i$ band. This study uses PTF $r$-band and ZTF $g$-band data. ASAS-SN initially operates in the $V$ band and later switches to the $g$ band. ATLAS observes with two broad optical filters: cyan ($c$) and orange ($o$). For MIR coverage, we use WISE and NEOWISE observations in the $W1$ (3.4 $\mu$m) and $W2$ (4.6 $\mu$m) bands, which help trace hot dust emission and constrain its spatial scales.

\begin{figure*}[!ht]
\centering 
\includegraphics[scale=0.7]{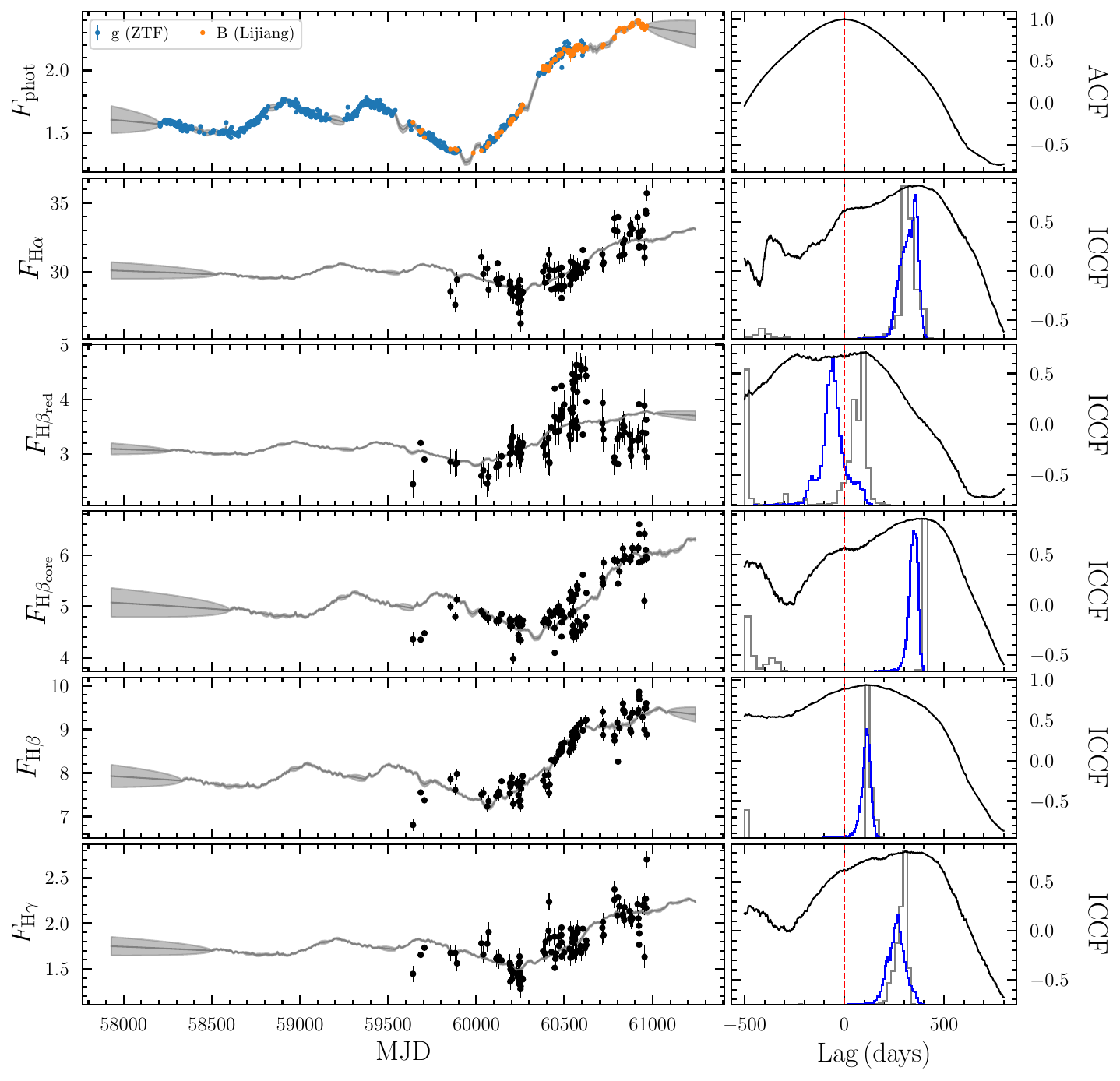}
\caption{Light curve and cross-correlation analysis results of emission lines. The left panels show the light curves. From top to bottom, they correspond to the photometric continuum, \ha, \hbred, \hbcor, \hb, and \hg. The gray solid lines and shaded regions represent the light curves reconstructed by JAVELIN, along with their associated uncertainties. Emission-line fluxes are measured in units of $10^{-13}~\ergscm$, while the photometric continuum is in arbitrary flux units. The right panels display the results of the ACF and ICCFs (black lines), and CCCDs (blue step lines). The gray step curves show the posterior JAVELIN time-lag distributions. The red dashed lines mark zero time lag. 
}
\label{fig:1}
\end{figure*}

\subsection{Intercalibration}\label{sec:2.4}
Our dataset includes photometric light curves from multiple surveys and instruments, each with distinct photometric systems and calibration methods. To ensure consistency across these heterogeneous data, we performed an intercalibration to bring all light curves onto a common flux scale. We adopt two intercalibration strategies: the first aligns only the ZTF $g$-band and Lijiang $B$-band data for measuring the time delay between the optical continuum and broad emission lines; the second incorporates additional archival data from CRTS, PTF, ASAS-SN, and ATLAS to study long-term optical-infrared variability.

Both intercalibration strategies are executed using PyCALI \citep{Li2014, Pycali2024}, which models light curve variability through a damped random walk (DRW) process. PyCALI simultaneously fits a multiplicative factor and an additive offset to align each light curve with a designated reference. We adopt the default parameter settings of PyCALI, and use the Lijiang $B$-band light curve as the reference for all cases. For the long-term variability analysis, we bin the intercalibrated optical data in 30-day intervals and the WISE data in 6-month intervals to improve the S/N and facilitate correlation studies. The final light curves used for time-delay measurements are shown in Figure~\ref{fig:1}, with the corresponding data listed in Table~\ref{tab:1}. The full multi-band light curves over the long-term baseline are presented in Figure~\ref{fig:a}.

\begin{figure*}[!ht]
\centering 
\includegraphics[scale=0.6]{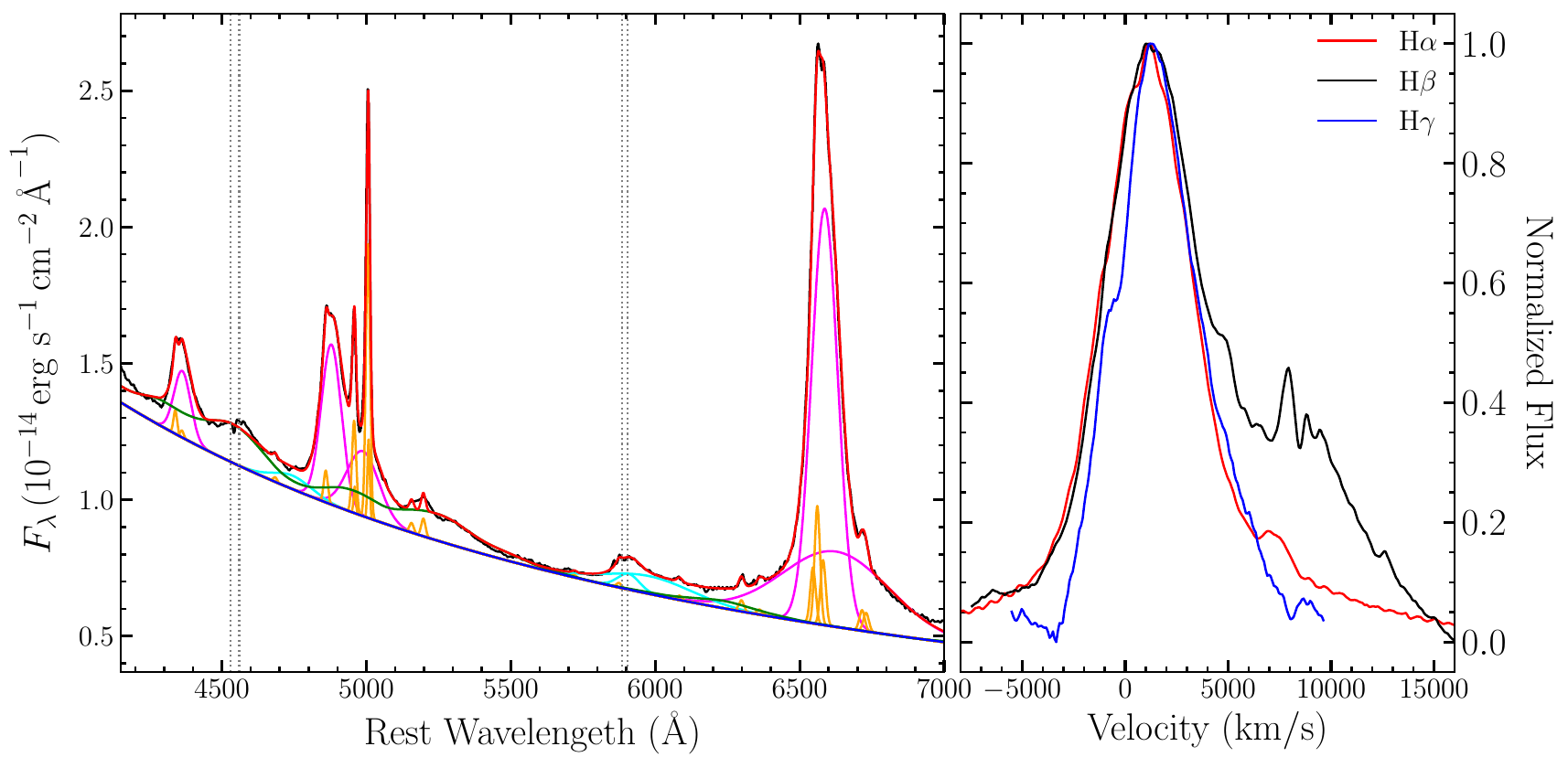}
\caption{Spectral fitting result for the mean spectrum. The left panel displays the observed spectrum (black line) and the best-fit model (red line), with individual components color-coded as follows: AGN continuum (blue), \feii\ template (green), broad Balmer lines (magenta), narrow lines (orange), and broad \heii\ and \hei\ lines (cyan). Vertical dotted lines indicate the positions of two telluric absorption bands that were masked during the fitting. The right panel presents the residual profiles of the broad \ha, \hb, and \hg\ lines after subtraction of the other best-fitting spectral components.
}
\label{fig:2}
\end{figure*}

\section{Analysis and Results} \label{sec:3}
\subsection{Spectral Fitting} \label{sec:3.1}
We extract the emission-line and continuum components through spectral fitting to properly separate blended features such as \feii. To verify the robustness of our results, we also apply the traditional integration method and find consistent variations between light curves. Our fitting procedure generally follows \citet{Feng2025a}, with a few modifications. First, the absence of detectable host galaxy features in our spectra eliminates the need for stellar templates. Second, since the observations use multiple grisms and slit widths, resulting in varying spectral resolutions, we construct separate mean spectra for each configuration. These mean spectra are used to estimate the instrumental broadening from the narrow-line widths and to constrain the individual spectral fits by fixing the widths and velocity shifts of \feii\ and \heii, as well as the narrow-line flux ratios relative to \oiii\ $\lambda$5007. Third, the fitting window is set to 4170-6960 \AA\ in the rest frame, covering \ha\ to \hg, with regions around 4543 \AA\ and 5895 \AA\ masked due to telluric absorption.

All spectra are corrected for Galactic extinction \citep{Fitzpatrick1999, Schlafly2011} and redshift prior to fitting. The fitting model includes: (1) a power-law continuum; (2) an \feii\ template from \citet{Boroson1992}; (3) double-Gaussian profiles for the broad \ha, \hb, and \hei\ $\lambda$5876, and single-Gaussian profiles for broad \hg\ and \heii\ $\lambda$4686; (4) double-Gaussian profiles for \oiii\ $\lambda\lambda$4959, 5007, and single Gaussians for all other narrow lines. The reported flux uncertainties include both Poisson noise and systematic errors, estimated following \citet{Li2022}. Figure \ref{fig:1} shows the resulting light curves, and the corresponding data are listed in Table \ref{tab:1}. Figure \ref{fig:2} presents representative spectral fits.

\begin{deluxetable*}{lccccccccccc}[!htbp]
 \tablecolumns{11}
\tablewidth{\textwidth}
\tabletypesize{\scriptsize}
\tablecaption{Light Curves}
\label{tab:1}
\tablehead{\multicolumn{6}{c}{Spectra}        &
      \colhead{}                         &
      \multicolumn{3}{c}{Photometry}     \\ 
      \cline{1-6} \cline{8-10}  
      \colhead{JD - 2458000}                       &
      \colhead{$F_{\rm H\alpha}$}               &
      \colhead{\hbred}         &
      \colhead{\hbcor}         &
      \colhead{$F_{\rm H\beta}$}         &
      \colhead{$F_{\rm H\gamma}$}               &
      \colhead{}                         &
      \colhead{JD - 2458000}     &
      \colhead{$F_{\rm phot}$}            &
      \colhead{Obs}
      }
\startdata
1641.44 & $32.07 \pm 0.67$ & $2.45 \pm 0.25$ & $4.36 \pm 0.13$ & $6.81 \pm 0.14$ & $1.44 \pm 0.09$ & & 206.37 & $1.575 \pm 0.006$ & ZTF\\ 
1685.34 & $33.85 \pm 0.67$ & $3.20 \pm 0.28$ & $4.35 \pm 0.17$ & $7.56 \pm 0.14$ & $1.65 \pm 0.09$ & & 207.59 & $1.555 \pm 0.006$ & ZTF\\ 
1706.27 & $32.74 \pm 0.66$ & $2.90 \pm 0.25$ & $4.48 \pm 0.12$ & $7.38 \pm 0.13$ & $1.73 \pm 0.09$ & & 207.96 & $1.566 \pm 0.011$ & ZTF\\ 
1854.12 & $28.55 \pm 0.66$ & $2.86 \pm 0.24$ & $5.00 \pm 0.09$ & $7.86 \pm 0.13$ & $1.67 \pm 0.09$ & & 209.01 & $1.560 \pm 0.011$ & ZTF\\ 
1881.01 & $27.59 \pm 0.66$ & $2.81 \pm 0.24$ & $4.80 \pm 0.09$ & $7.62 \pm 0.13$ & $1.67 \pm 0.09$ & & 210.49 & $1.575 \pm 0.008$ & ZTF\\ 
\enddata
\tablecomments{The emission-line flux is measured in units of $10^{-13}~\ergscm$. $F_{\rm phot}$ is given in arbitrary flux units. Obs includes data from ZTF ($g$-band) and Lijiang ($B$-band). 
\\
(This table is available in a machine-readable form in the online journal.)}
\end{deluxetable*}

\begin{deluxetable*}{lcccccccccccc}[!ht]
 \tablecolumns{12}
\tablewidth{\textwidth}
\tabletypesize{\scriptsize}
\tablecaption{Time Lags, Line Widths, and Black Hole Masses \label{tab:2}}	
\tablewidth{\textwidth}
\tablehead{ &&&& \multicolumn{2}{c}{Mean} &&  \multicolumn{2}{c}{rms}  \\
\cline{5-6} \cline{8-9} 
 \colhead{Line} & \colhead{$r_{\rm max}$} & \colhead{$\tau_{\rm cent} $} &\colhead{$\tau_{\rm JAV} $} &  \colhead{$\rm FWHM$} & \colhead{$\sigma_{\rm line}$} &&   \colhead{$\rm FWHM$} & \colhead{$\sigma_{\rm line}$} &
 \colhead{$M_{\rm VP}$}  & \colhead{$\mathdotM$} \\
&& \multicolumn{2}{c}{(days)} & \multicolumn{2}{c}{(\kms)} &&\multicolumn{2}{c}{(\kms)} &  \multicolumn{1}{c}{($\times 10^9 M_{\odot}$)}
   }		
\startdata
\ha & 0.87 &  $262.2_{-46.0}^{+18.8}$ &  $244.0_{-22.7}^{+32.6}$ & $5378 \pm 7 $ & $5703 \pm 16 $ && $4338 \pm 62 $ & $8297 \pm 202 $ &  $0.96_{-0.17}^{+0.07}$  &  $5.50_{-2.30}^{+1.48}$ \\ 
\hbred & 0.71 &  $-49.0_{-34.5}^{+50.5}$ &  $40.2_{-417.8}^{+34.5}$ & $8280 \pm 73 $ & $3631 \pm 36 $ && $7450 \pm 150 $ & $5503 \pm 42 $ &  --  &  -- \\ 
\hbcor & 0.86 &  $267.0_{-17.6}^{+16.6}$ &  $303.4_{-674.2}^{+7.1}$ & $4973 \pm 9 $ & $2114 \pm 3 $ && $4937 \pm 78 $ & $2224 \pm 25 $ &  $1.27_{-0.09}^{+0.09}$  &  $3.16_{-0.84}^{+0.82}$ \\ 
\hb & 0.94 &  $83.2_{-18.7}^{+17.5}$ &  $94.7_{-5.6}^{+8.6}$ & $5838 \pm 21 $ & $4283 \pm 11 $ && $4804 \pm 67 $ & $5320 \pm 78 $ &  $0.37_{-0.08}^{+0.08}$  &  $36.29_{-18.25}^{+17.36}$ \\ 
\hg & 0.81 &  $203.9_{-36.5}^{+31.0}$ &  $224.8_{-27.1}^{+14.8}$ & $4652 \pm 17 $ & $1977 \pm 7 $ && $5637 \pm 187 $ & $2421 \pm 36 $ &  $1.26_{-0.24}^{+0.21}$  &  $3.19_{-1.41}^{+1.27}$ \\ 
\hline
W1 & 0.92 &  $1565.2_{-271.1}^{+283.5}$ &  $1348.3_{-6.3}^{+280.2}$ &  --  &  -- &&  --  &  -- &  --  &  -- \\ 
W2 & 0.92 &  $2072.3_{-314.2}^{+274.8}$ &  $2008.9_{-377.5}^{+23.6}$  &  --  &  -- &&  --  &  -- &  --  &  -- \\
\enddata
\tablecomments{The time lags are measured in the rest frame. $\tau_{\rm cent}$ and $\tau_{\rm JAV}$ represent the time lags derived from the ICCF and JAVELIN methods, respectively. Line widths are corrected for instrumental broadening. The virial product, $M_{\rm VP}$, is calculated using the FWHM measured from the mean spectrum and the time lag $\tau_{\rm cent}$.
}
\end{deluxetable*}

\subsection{Time Lag Measurements} \label{sec:3.2}
We measure time lags between the optical continuum and the emission-line/MIR light curves using the interpolated cross-correlation function \citep[ICCF;][]{Gaskell1987}, the standard technique in time-series analysis. The lag is defined as the centroid of the CCF above 80\% of the peak correlation coefficient ($r \geq 0.8 r_{\rm max}$). Uncertainties are estimated via Monte Carlo simulations based on flux randomization and random subset selection (FR/RSS) \citep{Peterson1998}, which produce a cross-correlation centroid distribution (CCCD). We adopt the 15.87\% and 84.13\% of the CCCD as the lower and upper bounds of the lag uncertainty, respectively. 

For comparison, we also apply JAVELIN \citep{Zu2011}, which models variability as a DRW and assumes a top-hat transfer function. The lag is taken as the median of the posterior distribution, with the 68\% credible interval representing the uncertainty range. Both methods yield consistent results, and we adopt the ICCF-derived lags for subsequent analysis, given its widespread use and compatibility with previous RM studies.

Figures \ref{fig:1} and \ref{fig:a} present the autocorrelation functions (ACFs), CCFs, CCCDs, and JAVELIN-reconstructed light curves with model uncertainties. The measured lags and corresponding $r_{\rm max}$ values are summarized in Table~\ref{tab:2}.

\begin{figure}[!ht]
\centering 
\includegraphics[scale=0.6]{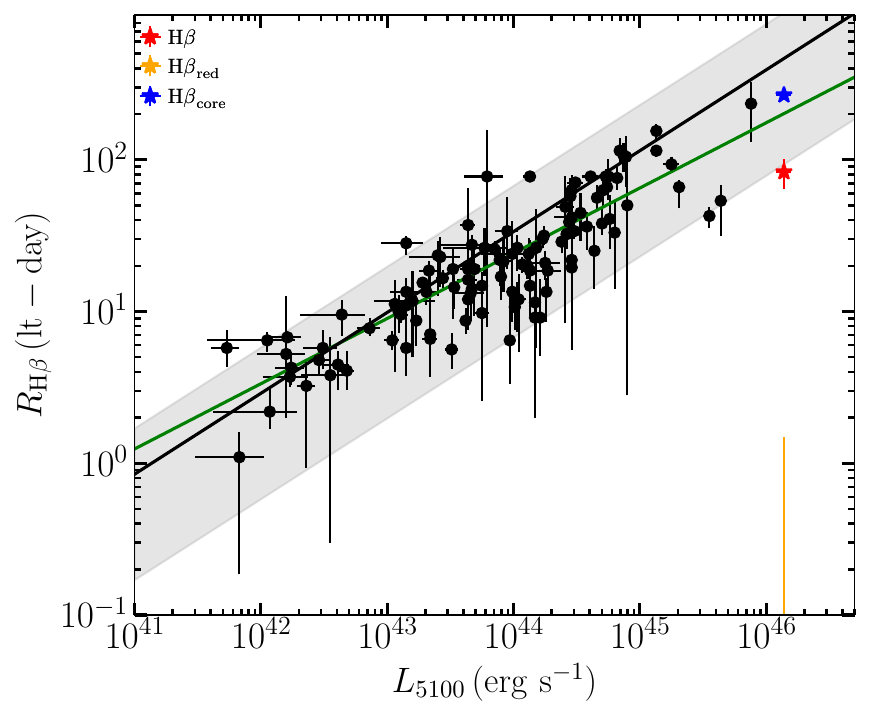}
\caption{The $R_{\rm H\beta}-L_{\rm 5100}$ relation. The black line shows the best-fit relation from \citet{Bentz2013}, while the shaded region marks the range between 0.2 and 2 times the radius predicted by this relation. Circles represent data points from the sample of \citet{Wang2024}, and the green line denotes their best-fit result. Colored stars show our measurements for E1821+643: the red star corresponds to the result from the total \hb\ profile, the orange star to the redshifted component (\hbred), and the blue star to the core component (\hbcor).
}
\label{fig:3}
\end{figure}

\subsection{Comparison with the $R_{\rm BLR}-L_{\rm 5100}$ Relation} \label{sec:3.3}
To our knowledge, E1821+643 is the most luminous AGN with a successful \hb\ RM measurement to date \citep{Wang2024, Woo2024}, providing a valuable opportunity to probe the $R_{\rm BLR}$-$L_{5100}$ relation at the high-luminosity end. Using the best-fitting power-law continuum at 5100 \AA\ from individual spectra (Section \ref{sec:3.1}), we derive a rest-frame monochromatic luminosity of $L_{5100} = (1.4 \pm 0.2) \times 10^{46} \ergs$. According to the $R_{\rm BLR}-L_{\rm 5100}$ relation from \citet{Bentz2013}, the expected \hb\ lag is 467.9 days, which is 5.6 times longer than our measured value of 83.2 days.

To test whether this deviation is unique to high-luminosity AGNs, we compare our result with the current \hb\ RM sample. We adopt the recent compilation by \citet{Wang2024}, which includes nearly all published \hb\ RM data to date. They employe a set of quantitative criteria and visual inspection to exclude unreliable lag measurements, ensuring high-quality results. Following \citet{Du2019}, we use averaged values for sources with multiple RM campaigns, resulting in a final sample of 112 AGNs.

We then arbitrarily shift the \citet{Bentz2013} relation downward by a factor of 5 (i.e., to 0.2$R_{\rm BLR}$) and find that, across the full luminosity range ($10^{43}$-$10^{46} \ergs$), nearly all the shortest RM-measured lags lie at this boundary. The upper boundary appears to be around 2$R_{\rm BLR}$. While we do not perform a formal statistical analysis of the scatter, this empirical envelope (0.2-2$R_{\rm BLR}$) encompasses 108 out of the 112 sources, with only 4 outliers exhibiting large uncertainties or modest deviations (Figure \ref{fig:3}).

This suggests that, for individual sources, the uncertainty in lags predicted from the $R_{\rm BLR}-L_{\rm 5100}$ relation can reach up to 1 order of magnitude. \citet{Dalla2020} reported that replacing $L_{5100}$ with the \hb\ luminosity can effectively mitigate host galaxy contamination without loss of precision. We therefore also examine the $R_{\rm BLR}-L_{\rm H\beta}$ relation and find that E1821+643 remains one of the most significant outliers, and the $\sim$1 dex envelope persisting in this parameter space as well.

\citet{Du2019} proposed that the deviation in high-accretion AGNs can be corrected using the parameter \Rfe, defined as the flux ratio of \feii\ to \hb. For E1821+643, with \Rfe $\sim$0.34, their method yields an expected lag of 313.3 days. While this prediction is shorter than that from \citet{Bentz2013}, it still exceeds our measured value by a factor of $\sim$3.8. We also compare our result with the relation given by \citet{Wang2024}, who performed a joint fit to both high- and low-accretion AGNs. Their updated relation predicts a lag of 201.5 days for E1821+643, which is closer to our measured core lag but still $\sim$2.4 times longer than the entire \hb. These comparisons suggest that, even after accounting for accretion-related effects, E1821+643 exhibits an anomalously short \hb\ lag that cannot be fully explained by current empirical relations.

\subsection{Black Hole Mass and Accretion Rate} \label{sec:3.4}
According to the virial relation described in Equation \ref{equ1}, we estimate the black hole mass for each broad emission line with a measured time lag and line width. The line widths are obtained from 1000 bootstrap realizations of the mean and rms spectra, constructed from the best-fit broad-line profiles in individual spectra. For each iteration, we randomly select $N$ spectra with replacement, remove duplicates, and compute new mean and rms spectra. The full width at half maximum (FWHM) and the line dispersion ($\sigma_{\rm line}$) are measured from each realization, with the final values and uncertainties taken as the mean and standard deviation of the resulting distributions. All line widths were corrected for instrumental broadening based on \oiii$\lambda$5007 line comparisons.

Although both FWHM and $\sigma_{\rm line}$ are measured, we adopt the FWHM from the mean spectrum as the velocity scale for black hole mass estimation to ensure consistency with previous studies. Here we adopt $f = 1$, following \citet{Du2019}, which is close to the average value of $f = 1.12$ reported by \citet{Woo2015}. This choice allows for direct comparison with previous results. Under this assumption, the black hole mass is equivalent to the virial product (i.e., $\mbh = M_{\rm VP}$). The resulting virial products for all broad lines are listed in Table \ref{tab:2}.

\citet{Shapovalova2016} also reported a black hole mass of $\mbh = 2.6 \times 10^9M_\odot$ based on the FWHM of the \hb\ line measured from the rms spectrum, adopting a significantly larger virial factor of $f = 5.5$, which corresponds to a virial product of $M_{\rm VP} = 4.7 \times 10^8M_\odot$. Recalculating the virial product using the FWHM from our rms spectrum yields $M_{\rm VP} = (4.10^{+0.74}_{-0.73}) \times 10^{8} \sunm$, in good agreement with their result.

We estimate the dimensionless accretion rate, $\mathdotM$, following the prescription of \citet{Du2019}:
\begin{equation}\label{equ2}
\mathdotM = 20.1\,\left(\frac{\ell_{44}}{\cos i}\right)^{3/2}m_7^{-2},
\end{equation}
where $\ell_{44}=L_{5100}/10^{44} \ergs$ is the monochromatic luminosity at 5100 \AA\ normalized to $10^{44} \ergs$, and $m_7 = M_{\bullet}/10^7\sunm$ is the black hole mass in units of $10^7 \sunm$.  The inclination angle of the accretion disk is denoted by $i$, and we adopt $\cos i = 0.75$, consistent with the assumption in \citet{Du2019}.

Using the continuum luminosity $L_{5100}$ determined in Section \ref{sec:3.3} and the black hole masses derived from each broad emission line, we calculate the corresponding values of $\mathdotM$. The results are summarized in Table~\ref{tab:2}. Due to the varying kinematic properties and emission regions of different broad lines, the black hole mass estimates--and consequently the derived accretion rates--exhibit significant variation. Given these differences, and considering that \hb\ is the most extensively monitored line in RM campaigns, we focus our analysis on \hb. This choice is further justified by the fact that other broad lines in our spectra are either weaker or located near the spectral edges, where the data quality is lower.

\begin{figure}[!ht]
\centering 
\includegraphics[scale=0.38]{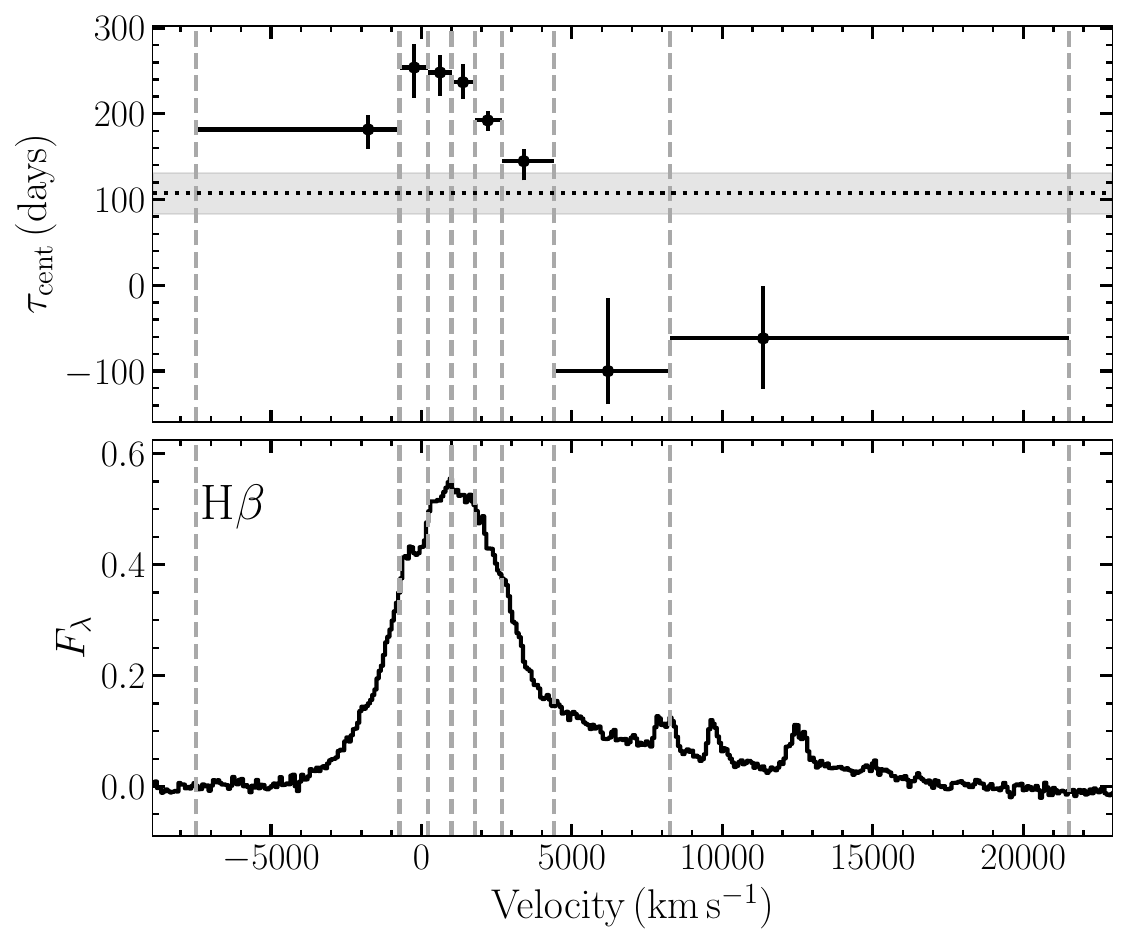}
\includegraphics[scale=0.38]{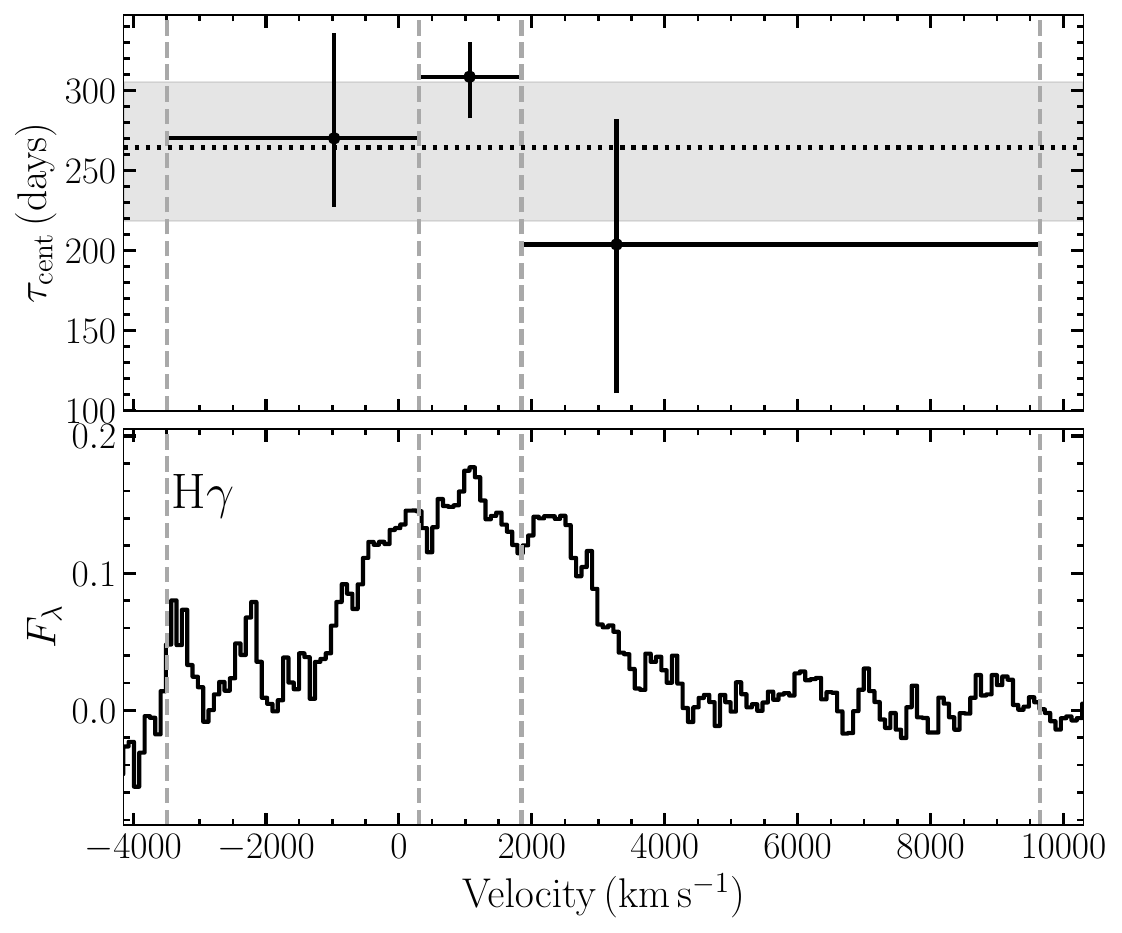}
\includegraphics[scale=0.38]{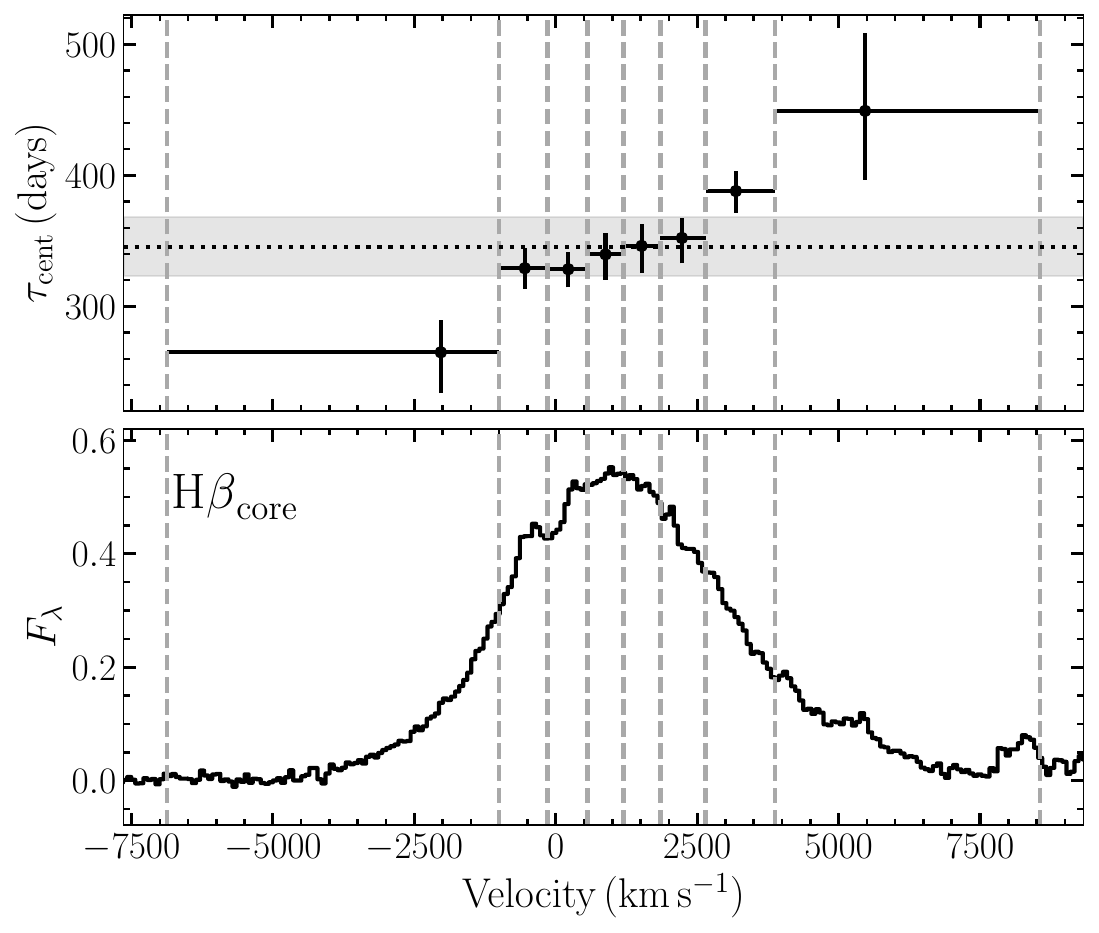}
\caption{Velocity-resolved time lag measurements for the broad \hb\ and \hg\ emission lines, as well as for the core component of \hb. In each top panel, the time lag as a function of velocity bin is shown. The horizontal dotted line indicates the mean time lags of the corresponding emission line, and the gray shaded region denotes the associated uncertainty range. In each bottom panel, the step line represents the residual rms spectrum, with vertical dashed lines marking the boundaries of the velocity bins used in the lag analysis. The flux is in $10^{-15}~\ergscma$.}
\label{fig:4}
\end{figure}

\subsection{Velocity-resolved Time Lags} \label{sec:3.5}
Velocity-resolved time lag analysis provides valuable insights into the geometry and kinematics of the BLR \citep[e.g.,][]{Stone2025}. The fundamental principle underlying this technique is that BLR clouds at different locations respond to continuum variations with distinct time delays and exhibit different line-of-sight velocities.

To perform this analysis, we first isolate the broad emission-line components from the observed spectra by subtracting all other spectral components, using the best-fit models described in Section~\ref{sec:3.1}. This process yields residual spectra that contain only the broad-line emission. From the residuals, we construct the rms spectrum for each broad line. Each rms profile is then divided into several velocity bins, with bin boundaries defined to ensure that each bin contains equal flux. Using these bin definitions, we extract light curves for each velocity bin from the residual spectra and measure the corresponding time lags and uncertainties using the ICCF method.

Among all the lines, \hb\ provides the most reliable results due to its high S/N and optimal position in the spectral range. In contrast, the \hei, \heii, and \hg\ lines are significantly weaker in our spectra, and the \ha\ line lies near the red edge of the spectral coverage, where reduced sensitivity and telluric absorption affect the data quality. As a result, velocity-resolved measurements for these lines are less reliable. Nevertheless, by applying a coarse binning scheme to \hg\ to improve the S/N ratio per bin, we obtain a lag profile that exhibits a similar trend to that of \hb, with longer lags in the blueshifted wing and shorter lags in the redshifted wing. Figure \ref{fig:4} presents the velocity-resolved lag measurements for \hb\ and \hg.

\section{Discussion} \label{sec:4}
\hb\ is the most commonly used emission line in RM studies, and the $R_{\rm BLR}$-$L_{5100}$ relation derived from \hb\ observations forms the foundation for single-epoch SMBH mass measurements across the universe. Our successful \hb\ RM campaign of the luminous quasar E1821+643, with $L_{5100} = (1.4 \pm 0.2) \times 10^{46} \ergs$, extends this relation to an unprecedented luminosity regime. Surprisingly, the measured \hb\ lag of 83.2 days is a factor of $\sim$5 shorter than the expected value of 467.9 days from the extrapolated $R_{\rm BLR}$-$L_{5100}$ relation \citep{Bentz2013}. This level of deviation is not unique to this object. When compared with the full \hb\ RM sample \citep{Wang2024}, we find that the shortest measured lags at all luminosities consistently fall along a lower envelope of $0.2R_{\rm BLR}$, while the upper envelope remains at $2 R_{\rm BLR}$. This $\sim$1 dex scatter significantly exceeds the widely quoted statistical scatter of 0.13-0.3 dex \citep[e.g.,][]{Bentz2013, Du2019, Wang2024}, and suggests that significant systematic uncertainties may be present in black hole mass estimates based on single-epoch spectroscopy. Below, we investigate the origin of this deviation and its implications for BLR structure and the reliability of SMBH mass measurements.

\subsection{The Origin of the Short Lag}
At moderate to low luminosities, \citet{Du2018} reported that AGNs with high accretion rates ($\mathdotM > 3$) tend to exhibit shorter reverberation lags than predicted by the canonical $R_{\rm BLR}$-$L_{5100}$ relation. E1821+643, with $\mathdotM = 36.29$ based on \hb\ measurements, demonstrates that the same mechanism may operate in the high-luminosity regime. However, the origin of this phenomenon remains unclear.

\citet{Wang2014} proposed a theoretical framework involving two BLR zones. In their model, when the accretion disk enters a super-Eddington state, the resulting slim disk anisotropically shields UV photons that ionize the BLR, altering the ionization state of one component and consequently changing the observed lag. While this model explains a reduction by a factor of $\sim$2, it cannot account for the factor-of-5 discrepancy observed in E1821+643.

Our spectral analysis reveals that the broad \hb\ profile in E1821+643 comprises two distinct components: a core component (\hbcor) centered near the systemic velocity with velocity shift of $1096 \pm 4$ \kms, and a highly redshifted tail (\hbred) with velocity shift of $7800 \pm 19$ \kms. We independently measure the time lags of these two components and find dramatically different results. The \hbcor\ exhibits a lag of $267.0_{-17.6}^{+16.6}$ days, much closer to the $R_{\rm BLR}$-$L_{5100}$ prediction, while the \hbred\ shows a lag of $-49.0_{-34.5}^{+50.5}$ days. This near-zero or even potentially negative lag suggests that the emitting region of \hbred\ is comparable to that of the optical continuum. The contribution from the short-lag redshifted tail reduces the flux/response-weighted mean lag of the entire \hb, explaining the observed deviation from the empirical relation.

The high redshift velocity and short lag of \hbred\ indicate emission from regions very close to the central SMBH. Such compact BLR structures are often modeled as a rapidly rotating disk ring  \citep{Chen1989}, which naturally produces a broad, double-peaked profile. Due to relativistic Doppler boosting, the blueshifted peak is typically stronger than the red side. RM studies have recently revealed possible signatures of such inner disk-like substructures in other AGNs, such as PG0026+129 \citep{Hu2020} and SDSSJ125809.31+351943.0 \citep{Nagoshi2024}. However, no prominent blueshifted component is observed in the \hb\ profile of E1821+643. This asymmetry suggests that the blue-emitting region lies at larger radii, implying an elliptical geometry for the inner BLR.

Our two-Gaussian decomposition is empirical rather than physically motivated. To test this hypothesis, we apply the elliptical disk model of \citet{Eracleous1995} to fit the emission-line profiles. Based on the above properties, we expect the position angle to be $\sim$270$^\circ$ in their formalism. As shown in Appendix~\ref{sec:AppxB}, this model successfully reproduces the observed line profiles, supporting our interpretation of an elliptical inner BLR structure.

Alternatively, if Doppler boosting is neglected, emission from regions close to the SMBH would be subject to strong gravitational redshift, producing a pronounced redshifted tail together with a stronger peak near zero velocity---features that closely resemble the entire \hb\ profile of E1821+643 \citep[see Figure 4 of][]{Corbin1997}. This characteristic has also been used to independently estimate black hole masses, which in turn enables studies of the impact of radiation pressure on the virial factor $f$ \citep{Liu2017, Liu2022, Liu2024}. Despite the apparent similarity in line profiles, emission from such gravitationally dominated regions is expected to exhibit a short lag for the line center, comparable to that of \hbred. This is inconsistent with the much longer lag observed for \hbcor. 

Another plausible explanation for the redshifted tail involves contamination from helium emission lines. \citet{Veron2002} proposed that \hei\ $\lambda4922$ and $\lambda5016$ emission can produce a spurious “red shelf” on the \hb\ profile. These \hei\ lines are expected to originate from regions closer to the ionizing source and thus naturally exhibit shorter time lags, consistent with that of the \hbred\ component. This scenario may be supported by a comparative analysis of the broad \ha, \hb, and \hg\ line profiles (the right panel of Figure \ref{fig:2}), which should reveal the strongest redward excess in \hb, followed by \ha, due to contribution from \hei\ $\lambda6678$. Unfortunately, the current low spectral resolution and insufficient S/N in the \ha\ and \hg\ regions limit further verification.

Regardless of the physical origin of \hbred, neglecting the lag between the UV and optical continua \citep{Zhou2025, Feng2025b} can introduce a considerable bias in RM measurements.
The UV-optical lag should have minimal impact on the outer \hbcor\ component, but could significantly affect the inner redshifted component, and ultimately reducing the flux/response-weighted mean lag. This interpretation can be tested observationally. MIR lags, such as those observed in WISE bands, typically lag the optical continuum by a factor of $\sim$10 compared to the BLR size \citep{Chen2023, Sun2025, Tomar2025}. For E1821+643, we measure $W1$ and $W2$ lags of $\sim$1800 and $\sim$2100 days, respectively, which are approximately 7 times the \hbcor\ lag, consistent with this expectation (Appendix~\ref{sec:AppxA}). This agreement suggests that \hbcor\ represents the ``true" BLR lag, while the composite \hb\ lag is biased by the inner component.

We note that the \hbcor\ lag still appears shorter than expected. If the elliptical disk model is correct, its blueshifted emission underlying the \hb\ core, though weaker, should contribute to the response. Consequently, the true \hbcor\ lag may be longer than our current measurement, implying a larger SMBH mass and a lower accretion rate.

\subsection{Spatially Distinct Structures in the BLR}
Previous velocity-resolved RM campaigns have revealed a wide diversity of velocity-lag patterns among different AGNs \citep{Denney2009, Hu2020, Feng2021a, Feng2021b, Villafana2022, Bao2022, Yao2024, Wang2025}. Within the traditional RM perspective, these patterns are typically interpreted as signatures of distinct BLR kinematics: longer lags on the red wing compared to the blue wing are taken as evidence for outflowing gas, while the opposite trend indicates inflow, and symmetric profiles with the longest lags near line center are attributed to virialized motion \citep{Bentz2010}. However, our recent work suggests that such interpretations may be overly simplistic \citep{Li2024}. For example, an elliptical Keplerian disk can produce velocity-resolved patterns that mimic radial motions without requiring actual inflow or outflow. Moreover, the gas distribution and kinematic properties can vary significantly from the inner to outer BLR \citep{Feng2024, Feng2025a}. These considerations suggest that the core and redshifted components in E1821+643 may differ not only in scale but also in their kinematic properties.

To investigate this possibility, we perform velocity-resolved RM measurements for the two components separately, following the approach described in Section \ref{sec:3.5}. As shown in Figure \ref{fig:4}, the \hbcor\ component exhibits a pattern characteristic of outflowing gas, with longer lags on the red wing than on the blue. This is in contrast to the entire \hb\ profile, which shows a signature of inflow. Such inflow signatures are commonly observed in previous RM campaigns, and many of these objects also display redward asymmetries in their broad emission lines \citep[e.g.,][]{Feng2021a, Bao2022}. Notably, the redshifted velocities in those cases are typically more modest, and their contribution to the mean time lag is generally less significant than what is observed here. 

Due to its large uncertainties, the velocity-resolved lag profile of \hbred\ remains inconclusive. We also caution that the above interpretation is phenomenological in nature and does not rely on a physically motivated model. Nevertheless, the BLR in the object unambiguously harbors complex structure. These results underscore a critical caveat in interpreting velocity-resolved RM. When multiple emitting regions with different gas distributions and kinematics are blended together, the resulting velocity-lag profile may not represent the true motion of any single component. This complexity highlights the limitations of simple kinematic interpretations and calls for a more physically motivated decomposition of the line profile and time series data in future RM analyses.

\subsection{Impact on SMBH Mass Estimates}
Single-epoch mass estimates of SMBHs are based on two key assumptions: that the BLR follows virial motion as a coherent system, with its characteristic radius scaling tightly with luminosity. Generally, the largest source of uncertainty in such measurements, as well as in RM, is attributed to the virial factor $f$, which is caused by the unknown BLR geometry, kinematics, and inclination. The uncertainty in $f$ can span a factor of 2-6 \citep[e.g.,][]{Mejia-Restrepo2018}, and is typically calibrated using the \mbh-$\sigma_*$ relation \citep{Onken2004, Woo2015}. However, our finding of distinct BLR components with dramatically different spatial scales and kinematic properties in E1821+643 reveals a more substantial systematic uncertainty.

The coexistence of multiple BLR components introduces significant complexity in determining both the characteristic BLR radius and velocity. While \hbcor\ basically follows the expected $R_{\rm BLR}$-$L_{5100}$ relation, the mean lag is reduced due to contributions from the inner \hbred. This suggests that the scatter in the $R_{\rm BLR}$-$L_{5100}$ relation may reflect variations in the relative contributions of multiple BLR zones across the AGN population, which can introduce up to an order-of-magnitude uncertainty in the inferred BLR size. Moreover, the different velocity-resolved patterns between these components can also complicate the velocity measurements from single-epoch spectra. The combined effect can bias the virial mass estimate by factors of tens or more, far exceeding the traditionally quoted uncertainties.

A natural question arises: why is such a multi-component BLR structure detectable in E1821+643? Most previously studied high-accretion-rate AGNs are narrow-line Seyfert 1 galaxies, which typically exhibit relatively narrow and weak \hb\ profiles and strong \feii\ emission. These spectral characteristics can compromise the detection of redshifted tail. In contrast, E1821+643 exhibits a broad and strong \hb, with the redshifted tail extending beyond the \oiii$\lambda5007$ region. Additionally, the \feii\ emission in this object is smooth and weak, allowing the redshifted tail to emerge clearly. These favorable spectral conditions enable the detection of distinct BLR zones in this object. These findings suggest that multi-component BLR structures may be common in high-accretion-rate AGNs but remain hidden in most objects due to observational limitations. This hypothesis can be tested with future high-quality spectroscopic monitoring campaigns.

\section{Conclusion} \label{sec:5}
We present a four-year RM campaign of the luminous quasar E1821+643 using photometric and spectroscopic monitoring from the Lijiang 2.4-m telescope, supplemented with archival optical and MIR data. Our main findings are summarized as follows:

\begin{enumerate}
\item The measured \hb\ lag of $83.2_{-18.7}^{+17.5}$ days is a factor of 5.6 shorter than predicted by the canonical $R_{\rm BLR}$-$L_{5100}$ relation. By compiling the full \hb\ RM sample, we find that this significant deviation ($0.2R_{\rm BLR}$) surprisingly defines a lower envelope of measured lags across the entire luminosity range, whereas the upper envelope lies near $2R_{\rm BLR}$. This implies that the scatter in the $R_{\rm BLR}$-$L_{5100}$ relation can reach up to 1 dex, significantly exceeding the commonly quoted statistical uncertainties of 0.13-0.3 dex.

\item Spectral decomposition reveals two distinct components in the broad \hb\ profile: a core component (\hbcor) with a lag of $267.0_{-17.6}^{+16.6}$ days that is closer to the expectation of the $R_{\rm BLR}$-$L_{5100}$ relation, and an extremely redshifted tail (\hbred) with a much shorter lag of $-49.0_{-34.5}^{+50.5}$ days. The \hbred\ not only accounts for the shortened overall \hb\ lag, but also leads to misinterpretation of the BLR kinematics inferred from velocity-resolved lags. The combination of these effects can introduce systematic uncertainties in single-epoch black hole mass estimates by factors of tens.

\item Our findings provide critical insight into the origin of shortened reverberation lags in high-accretion-rate AGNs and reveal previously unrecognized systematic uncertainties in SMBH mass estimates. The presence of multiple BLR components with distinct spatial scales and kinematic properties challenges the fundamental assumptions underlying single-epoch mass estimators, and has profound implications for their reliability.
\end{enumerate}

Future high-quality RM observations will be essential to determine the prevalence of such multi-component BLR structures across the AGN population, to improve physical models of the line-emitting regions, and to refine the application of the $R_{\rm BLR}$-$L_{5100}$ relation in black hole mass determinations and cosmological studies.

\begin{acknowledgments}
We appreciate the anonymous referee for their helpful comments and suggestions. We also thank Xiaolin Yang for insightful discussions and assistance with the implementation of the elliptical disk model, and Yongjie Chen for discussions on the survey data. This work is supported by National Key R\&D Program of China (No. 2021YFA1600404 and 2023YFA1607903), the National Natural Science Foundation of China (Nos. 12203096, 12573021, 12303022, 12322303, and 12373018), Yunnan Fundamental Research Projects (grants NO. 202301AT070358 and 202301AT070339), and the science research grants from the China Manned Space Project with (Nos. CMS-CSST-2025-A02, CMS-CSST-2025-A07). 

We acknowledge the support of the staff of the Lijiang 2.4 m telescope. Funding for the telescope has been provided by Chinese Academy of Sciences and the People’s Government of Yunnan Province. 

This work has made use of data from the Asteroid Terrestrial-impact Last Alert System (ATLAS) project. The Asteroid Terrestrial-impact Last Alert System (ATLAS) project is primarily funded to search for near earth asteroids through NASA grants NN12AR55G, 80NSSC18K0284, and 80NSSC18K1575; byproducts of the NEO search include images and catalogs from the survey area. This work was partially funded by Kepler/K2 grant J1944/80NSSC19K0112 and HST GO-15889, and STFC grants ST/T000198/1 and ST/S006109/1. The ATLAS science products have been made possible through the contributions of the University of Hawaii Institute for Astronomy, the Queen’s University Belfast, the Space Telescope Science Institute, the South African Astronomical Observatory, and The Millennium Institute of Astrophysics (MAS), Chile.
\end{acknowledgments}

\begin{contribution}
H.-C.F. conceived and initiated this project. S.-S.L. and H.-C.F. contributed to the observations, data reduction, analysis, and manuscript preparation. J.-C.W. and H.-C.F. discussed the elliptical disk fitting for the broad emission line profile. All authors participated in the discussion and interpretation of the results.
\end{contribution}

\facility{YAO:2.4m}
\software{PyRAF \citep{Pyraf2012}, DASpec \citep{Du2024}, PyCALI \citep{Pycali2024}, JAVELIN \citep{Zu2011}.}


\appendix
\renewcommand{\thefigure}{\Alph{section}\arabic{figure}}
\setcounter{figure}{0}
\section{BLR-Torus Size Comparison} \label{sec:AppxA}
We utilize the intercalibrated long-term optical photometric light curves from Section \ref{sec:2.4} together with WISE $W1$ and $W2$ observations, and adopt the methodology described in Section \ref{sec:3.2} to measure the time lags between the optical and MIR emission. The results are presented in Table \ref{tab:2}, with the CCF and JAVELIN analyses yielding nearly identical values. The size ratio of $R_{\rm H\beta}:R_{W1}:R_{W2}$ is $1:18.8:24.9$, which significantly deviates from the typical values, e.g., \citet{Chen2023} and \citet{Tomar2025} found $1:9.2:11.2$ and $1:9:12$, respectively. This notable discrepancy suggests that the measured $R_{\rm H\beta}$ may not fully reflect the intrinsic size of the BLR.

However, when considering only the core component of the \hb\ emission, the inferred ratio becomes $1:5.9:7.8$, which is much closer to the expected values from previous studies. This result provides further evidence for our interpretation that the \hbcor\ traces the canonical BLR emission, while the redshifted component leads to an overall shortening of the measured \hb\ lag.

\begin{figure*}[!ht]
\centering 
\includegraphics[scale=0.7]{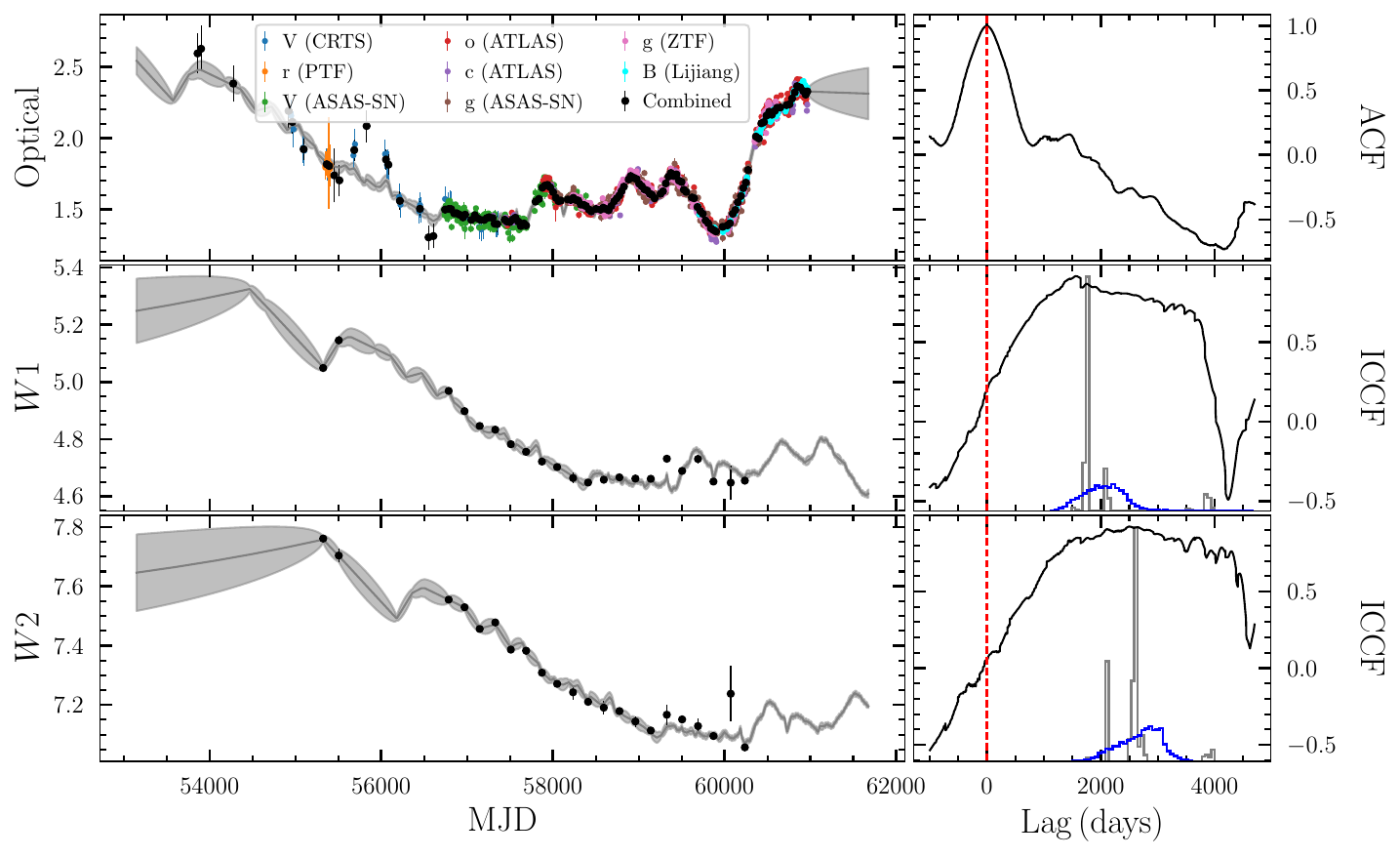}
\caption{Same as Figure \ref{fig:1}, but for the infrared data.
}
\label{fig:a}
\end{figure*}

\setcounter{figure}{0}
\section{Testing for a Possible Contribution from an Elliptical Disk in Broad \hb} \label{sec:AppxB}
To test whether the \hb\ emission line contains a contribution from a disk component, we perform an additional fit to the residual spectrum obtained by subtracting the \feii\ and continuum components, as described in Section \ref{sec:3.1}. The fitting is carried out over the wavelength range 4700-5100 \AA, using the same model components as in Section \ref{sec:3.1}, except that the broad \hb\ profile is modeled with a combination of a single Gaussian and an elliptical accretion disk component.

\begin{figure*}[!ht]
\centering 
\includegraphics[scale=0.45]{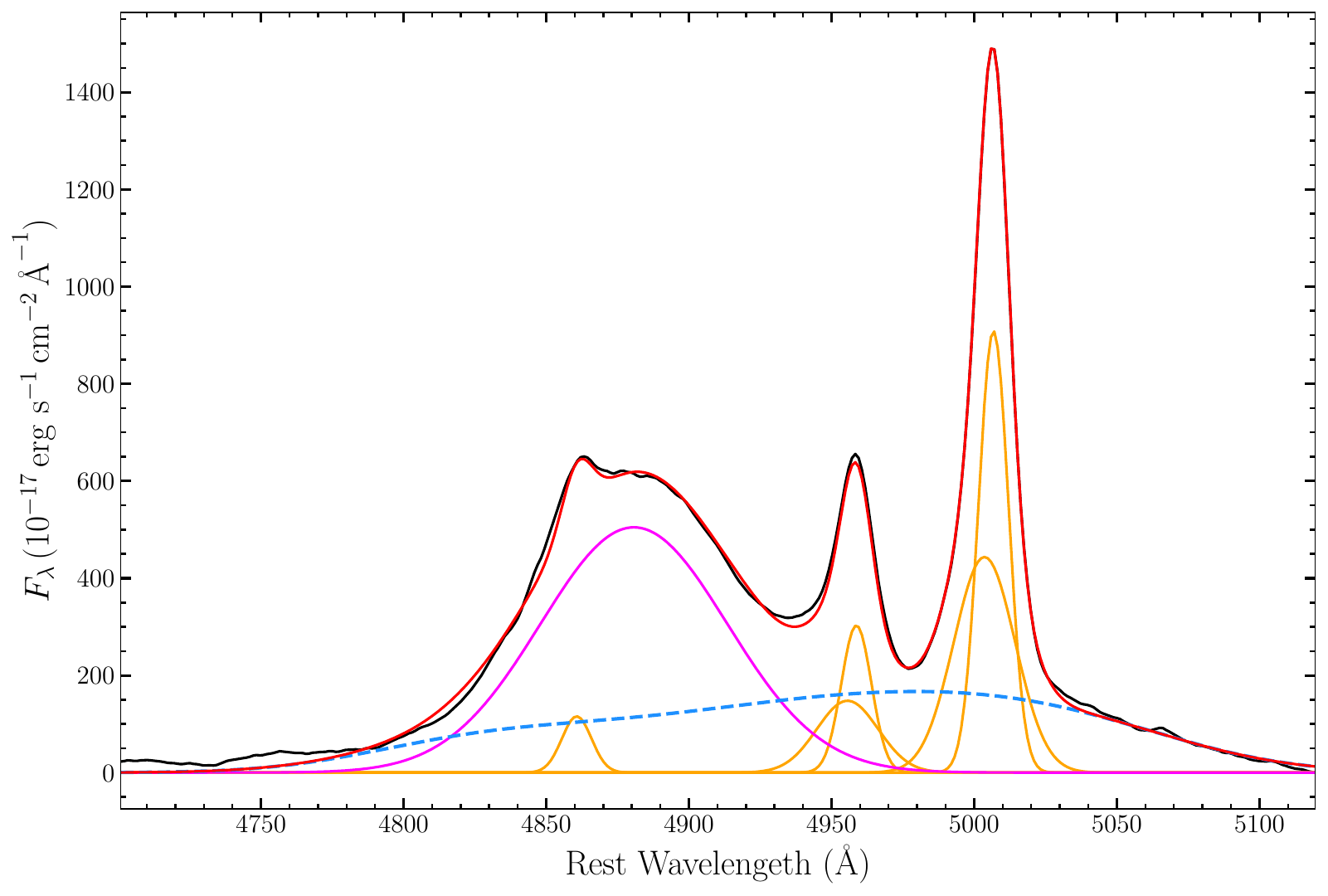}
\caption{Fitting result of the mean spectrum in the \hb\ region after subtracting the \feii\ and continuum components. The color scheme follows that of Figure \ref{fig:2}, with the disk component shown as a dashed light blue line.}
\label{fig:b}
\end{figure*}

The elliptical disk model follows the prescription of \citet{Eracleous1995}, with a minor modification: in their Equation (9), the sign of the $P^r$ component should depend on the azimuthal angle at which it is evaluated. This modification yields a more physically consistent treatment of the radial velocity component across the disk. The best-fit parameters for the disk component are: major-axis position angle $\phi_0 = 297.6^\circ$, ellipticity $e = 0.47$, inclination angle $i = 37.9^\circ$, inner pericenter radius $r_1 = 344.7 R_{\rm g}$, outer pericenter radius $r_2 = 3840.8 R_{\rm g}$, radial emissivity index $q = 3.5$, and local velocity dispersion $\sigma = 2200.0$ \kms. The fitting result is shown in Figure \ref{fig:b}.

{}


\begin{thebibliography}{}
\bibitem[Bao et al.(2022)]{Bao2022} Bao, D.-W., Brotherton, M.~S., Du, P., et al.\ 2022, \apjs, 262, 1, 14. doi:10.3847/1538-4365/ac7beb

\bibitem[Bentz et al.(2013)]{Bentz2013} Bentz, M.~C., Denney, K.~D., Grier, C.~J., et al.\ 2013, \apj, 767, 2, 149. doi:10.1088/0004-637X/767/2/149

\bibitem[Bentz et al.(2010)]{Bentz2010} Bentz, M.~C., Walsh, J.~L., Barth, A.~J., et al.\ 2010, \apj, 716, 2, 993. doi:10.1088/0004-637X/716/2/993

\bibitem[Blandford \& McKee(1982)]{Blandford1982} Blandford, R.~D. \& McKee, C.~F.\ 1982, \apj, 255, 419. doi:10.1086/159843

\bibitem[Boroson \& Green(1992)]{Boroson1992} Boroson, T.~A. \& Green, R.~F.\ 1992, \apjs, 80, 109. doi:10.1086/191661

\bibitem[Chen et al.(1989)]{Chen1989} Chen, K., Halpern, J.~P., \& Filippenko, A.~V.\ 1989, \apj, 339, 742. doi:10.1086/167332

\bibitem[Cao et al.(2025)]{Cao2025} Cao, S., Mandal, A.~K., Zaja{\v{c}}ek, M., et al.\ 2025, \prd, 111, 8, 083545. doi:10.1103/PhysRevD.111.083545

\bibitem[Chen et al.(2023)]{Chen2023} Chen, Y.-J., Liu, J.-R., Zhai, S., et al.\ 2023, \mnras, 522, 3, 3439. doi:10.1093/mnras/stad1136

\bibitem[Cho et al.(2023)]{Cho2023} Cho, H., Woo, J.-H., Wang, S., et al.\ 2023, \apj, 953, 2, 142. doi:10.3847/1538-4357/ace1e5

\bibitem[Corbin(1997)]{Corbin1997} Corbin, M.~R.\ 1997, \apj, 485, 2, 517. doi:10.1086/304474

\bibitem[Dalla Bont{\`a} et al.(2020)]{Dalla2020} Dalla Bont{\`a}, E., Peterson, B.~M., Bentz, M.~C., et al.\ 2020, \apj, 903, 2, 112. doi:10.3847/1538-4357/abbc1c

\bibitem[Denney et al.(2009)]{Denney2009} Denney, K.~D., Peterson, B.~M., Pogge, R.~W., et al.\ 2009, \apjl, 704, 2, L80. doi:10.1088/0004-637X/704/2/L80

\bibitem[Drake et al.(2009)]{Drake2009} Drake, A.~J., Djorgovski, S.~G., Mahabal, A., et al.\ 2009, \apj, 696, 1, 870. doi:10.1088/0004-637X/696/1/870

\bibitem[Du(2024)]{Du2024} Du, P.\ 2024, DASpec: A code for spectral decomposition of active galactic nuclei, v0.8, Zenodo, doi:10.5281/zenodo.12578528

\bibitem[Du et al.(2016)]{Du2016} Du, P., Lu, K.-X., Zhang, Z.-X., et al.\ 2016, \apj, 825, 2, 126. doi:10.3847/0004-637X/825/2/126

\bibitem[Du \& Wang(2019)]{Du2019} Du, P. \& Wang, J.-M.\ 2019, \apj, 886, 1, 42. doi:10.3847/1538-4357/ab4908

\bibitem[Du et al.(2018)]{Du2018} Du, P., Zhang, Z.-X., Wang, K., et al.\ 2018, \apj, 856, 1, 6. doi:10.3847/1538-4357/aaae6b

\bibitem[Eracleous et al.(1995)]{Eracleous1995} Eracleous, M., Livio, M., Halpern, J.~P., et al.\ 1995, \apj, 438, 610. doi:10.1086/175104

\bibitem[Feng et al.(2025a)]{Feng2025a} Feng, H.-C., Li, S.-S., Bai, J.~M., et al.\ 2025a, \apj, 979, 2, 131. doi:10.3847/1538-4357/ad9c71

\bibitem[Feng et al.(2025b)]{Feng2025b} Feng, H.-C., Li, S.-S., Sun, M., et al.\ 2025b, , arXiv:2512.18276. 

\bibitem[Feng et al.(2024)]{Feng2024} Feng, H.-C., Li, S.-S., Bai, J.~M., et al.\ 2024, \apj, 976, 2, 176. doi:10.3847/1538-4357/ad8568

\bibitem[Feng et al.(2021a)]{Feng2021a} Feng, H.-C., Hu, C., Li, S.-S., et al.\ 2021a, \apj, 909, 1, 18. doi:10.3847/1538-4357/abd851

\bibitem[Feng et al.(2021b)]{Feng2021b} Feng, H.-C., Liu, H.~T., Bai, J.~M., et al.\ 2021b, \apj, 912, 2, 92. doi:10.3847/1538-4357/abefe0

\bibitem[Feng et al.(2020)]{Feng2020} Feng, H.-C., Liu, H.~T., Bai, J.~M., et al.\ 2020, \apj, 888, 30. doi:10.3847/1538-4357/ab594b

\bibitem[Gaskell \& Peterson(1987)]{Gaskell1987} Gaskell, C.~M. \& Peterson, B.~M.\ 1987, \apjs, 65, 1. doi:10.1086/191216

\bibitem[Fitzpatrick(1999)]{Fitzpatrick1999} Fitzpatrick, E.~L.\ 1999, \pasp, 111, 63. doi:10.1086/316293

\bibitem[GRAVITY Collaboration et al.(2024)]{GRAVITY2024} GRAVITY Collaboration, Amorim, A., Bourdarot, G., et al.\ 2024, \aap, 684, A167. doi:10.1051/0004-6361/202348167

\bibitem[Hu et al.(2020)]{Hu2020} Hu, C., Li, S.-S., Guo, W.-J., et al.\ 2020, \apj, 905, 1, 75. doi:10.3847/1538-4357/abc2da

\bibitem[Kaspi et al.(2000)]{Kaspi2000} Kaspi, S., Smith, P.~S., Netzer, H., et al.\ 2000, \apj, 533, 2, 631. doi:10.1086/308704

\bibitem[Law et al.(2009)]{Law2009} Law, N.~M., Kulkarni, S.~R., Dekany, R.~G., et al.\ 2009, \pasp, 121, 886, 1395. doi:10.1086/648598

\bibitem[Li et al.(2025)]{Li2025} Li, S.-J., Ning, X.-W., Ma, Y.-S., et al.\ 2025, \apj, 988, 2, 273. doi:10.3847/1538-4357/adea4b

\bibitem[Li et al.(2024)]{Li2024} Li, S.-S., Feng, H.-C., Liu, H.~T., et al.\ 2024, \apj, 972, 1, 105. doi:10.3847/1538-4357/ad60c1

\bibitem[Li et al.(2022)]{Li2022} Li, S.-S., Feng, H.-C., Liu, H.~T., et al.\ 2022, \apj, 936, 75. doi:10.3847/1538-4357/ac8745

\bibitem[Li et al.(2021)]{Li2021} Li, S.-S., Yang, S., Yang, Z.-X., et al.\ 2021, \apj, 920, 9. doi:10.3847/1538-4357/ac116e

\bibitem[Li(2024)]{Pycali2024} Li, Y.-R.\ 2024, PyCALI: A Bayesian package for intercalibrating light curves, v0.2.3, Zenodo, doi:10.5281/zenodo.10700132

\bibitem[Li et al.(2014)]{Li2014} Li, Y.-R., Wang, J.-M., Hu, C., et al.\ 2014, \apjl, 786, 1, L6. doi:10.1088/2041-8205/786/1/L6

\bibitem[Li et al.(2018)]{Li2018} Li, Y.-R., Songsheng, Y.-Y., Qiu, J., et al.\ 2018, \apj, 869, 137. doi:10.3847/1538-4357/aaee6b

\bibitem[Liu et al.(2024)]{Liu2024} Liu, H.~T., Feng, H.-C., Li, S.-S., et al.\ 2024, \apj, 963, 1, 30. doi:10.3847/1538-4357/ad1ab8

\bibitem[Liu et al.(2022)]{Liu2022} Liu, H.~T., Feng, H.-C., Li, S.-S., et al.\ 2022, \apj, 928, 1, 60. doi:10.3847/1538-4357/ac559b

\bibitem[Liu et al.(2017)]{Liu2017} Liu, H.~T., Feng, H.~C., \& Bai, J.~M.\ 2017, \mnras, 466, 3, 3323. doi:10.1093/mnras/stw3261

\bibitem[Lu et al.(2022)]{Lu2022} Lu, K.-X., Bai, J.-M., Wang, J.-M., et al.\ 2022, \apjs, 263, 1, 10. doi:10.3847/1538-4365/ac94d3

\bibitem[Mainzer et al.(2011)]{Mainzer2011} Mainzer, A., Grav, T., Bauer, J., et al.\ 2011, \apj, 743, 2, 156. doi:10.1088/0004-637X/743/2/156

\bibitem[Masci et al.(2019)]{Masci2019} Masci, F.~J., Laher, R.~R., Rusholme, B., et al.\ 2019, \pasp, 131, 018003. doi:10.1088/1538-3873/aae8ac

\bibitem[Mej{\'\i}a-Restrepo et al.(2018)]{Mejia-Restrepo2018} Mej{\'\i}a-Restrepo, J.~E., Lira, P., Netzer, H., et al.\ 2018, Nature Astronomy, 2, 63. doi:10.1038/s41550-017-0305-z

\bibitem[Nagoshi et al.(2024)]{Nagoshi2024} Nagoshi, S., Iwamuro, F., Yamada, S., et al.\ 2024, \mnras, 529, 1, 393. doi:10.1093/mnras/stae319

\bibitem[Negrete et al.(2018)]{Negrete2018} Negrete, C.~A., Dultzin, D., Marziani, P., et al.\ 2018, \aap, 620, A118. doi:10.1051/0004-6361/201833285

\bibitem[Onken et al.(2004)]{Onken2004} Onken, C.~A., Ferrarese, L., Merritt, D., et al.\ 2004, \apj, 615, 645. doi:10.1086/424655

\bibitem[Panda \& {\'S}niegowska(2024)]{Panda2024} Panda, S. \& {\'S}niegowska, M.\ 2024, \apjs, 272, 1, 13. doi:10.3847/1538-4365/ad344f

\bibitem[Peterson et al.(1998)]{Peterson1998} Peterson, B.~M., Wanders, I., Horne, K., et al.\ 1998, \pasp, 110, 660. doi:10.1086/316177

\bibitem[Peterson(1993)]{Peterson1993} Peterson, B.~M.\ 1993, \pasp, 105, 247. doi:10.1086/133140

\bibitem[Planck Collaboration et al.(2020)]{PlanckCollaboration2020} Planck Collaboration, Aghanim, N., Akrami, Y., et al.\ 2020, \aap, 641, A6. doi:10.1051/0004-6361/201833910

\bibitem[Rakshit et al.(2020)]{Rakshit2020} Rakshit, S., Stalin, C.~S., \& Kotilainen, J.\ 2020, \apjs, 249, 1, 17. doi:10.3847/1538-4365/ab99c5

\bibitem[Schlafly \& Finkbeiner(2011)]{Schlafly2011} Schlafly, E.~F. \& Finkbeiner, D.~P.\ 2011, \apj, 737, 103. doi:10.1088/0004-637X/737/2/103

\bibitem[{{Science Software Branch at STScI}(2012)}]{Pyraf2012}
{Science Software Branch at STScI}. 2012, {PyRAF: Python alternative for IRAF}, Astrophysics Source Code Library, record ascl:1207.011

\bibitem[Shappee et al.(2014)]{Shappee2014} Shappee, B.~J., Prieto, J.~L., Grupe, D., et al.\ 2014, \apj, 788, 48. doi:10.1088/0004-637X/788/1/48

\bibitem[Shapovalova et al.(2016)]{Shapovalova2016} Shapovalova, A.~I., Popovi{\'c}, L. {\v{C}}., Chavushyan, V.~H., et al.\ 2016, \apjs, 222, 2, 25. doi:10.3847/0067-0049/222/2/25

\bibitem[Stone et al.(2025)]{Stone2025} Stone, Z., Shen, Y., Anderson, S.~F., et al.\ 2025, \apj, 991, 2, 218. doi:10.3847/1538-4357/adfd4c

\bibitem[Sun et al.(2025)]{Sun2025} Sun, J., Guo, H., Zuo, W., et al.\ 2025, \apjl, 989, 2, L26. doi:10.3847/2041-8213/adf3a3

\bibitem[Tomar et al.(2025)]{Tomar2025} Tomar, A., Rakshit, S., Mandal, A.~K., et al.\ 2025, \apj, 993, 2, 203. doi:10.3847/1538-4357/ae0f96

\bibitem[Tonry et al.(2018)]{Tonry2018} Tonry, J.~L., Denneau, L., Heinze, A.~N., et al.\ 2018, \pasp, 130, 988, 064505. doi:10.1088/1538-3873/aabadf

\bibitem[V{\'e}ron et al.(2002)]{Veron2002} V{\'e}ron, P., Gon{\c{c}}alves, A.~C., \& V{\'e}ron-Cetty, M.-P.\ 2002, \aap, 384, 826. doi:10.1051/0004-6361:20020072

\bibitem[Villafa{\~n}a et al.(2022)]{Villafana2022} Villafa{\~n}a, L., Williams, P.~R., Treu, T., et al.\ 2022, \apj, 930, 1, 52. doi:10.3847/1538-4357/ac6171

\bibitem[Wang et al.(2025)]{Wang2025} Wang, S., Woo, J.-H., Barth, A.~J., et al.\ 2025, \apj, 983, 1, 45. doi:10.3847/1538-4357/adbca5

\bibitem[Wang \& Woo(2024)]{Wang2024} Wang, S. \& Woo, J.-H.\ 2024, \apjs, 275, 1, 13. doi:10.3847/1538-4365/ad74f2

\bibitem[Wang et al.(2019)]{Wang2019} Wang, C.-J., Bai, J.-M., Fan, Y.-F., et al.\ 2019, Research in Astronomy and Astrophysics, 19, 10, 149. doi:10.1088/1674-4527/19/10/149

\bibitem[Wang et al.(2014)]{Wang2014} Wang, J.-M., Qiu, J., Du, P., et al.\ 2014, \apj, 797, 1, 65. doi:10.1088/0004-637X/797/1/65

\bibitem[Woo et al.(2024)]{Woo2024} Woo, J.-H., Wang, S., Rakshit, S., et al.\ 2024, \apj, 962, 1, 67. doi:10.3847/1538-4357/ad132f

\bibitem[Woo et al.(2015)]{Woo2015} Woo, J.-H., Yoon, Y., Park, S., et al.\ 2015, \apj, 801, 38. doi:10.1088/0004-637X/801/1/38

\bibitem[Wright et al.(2010)]{Wright2010} Wright, E.~L., Eisenhardt, P.~R.~M., Mainzer, A.~K., et al.\ 2010, \aj, 140, 6, 1868. doi:10.1088/0004-6256/140/6/1868

\bibitem[Wu et al.(2015)]{Wu2015} Wu, X.-B., Wang, F., Fan, X., et al.\ 2015, \nat, 518, 7540, 512. doi:10.1038/nature14241

\bibitem[Xin et al.(2020)]{Xin2020} Xin, Y.-X., Bai, J.-M., Lun, B.-L., et al.\ 2020, Research in Astronomy and Astrophysics, 20, 9, 149. doi:10.1088/1674-4527/20/9/149

\bibitem[Yang et al.(2021)]{Yang2021} Yang, J., Wang, F., Fan, X., et al.\ 2021, \apj, 923, 2, 262. doi:10.3847/1538-4357/ac2b32

\bibitem[Yao et al.(2024)]{Yao2024} Yao, Z.-H., Yang, S., Guo, W.-J., et al.\ 2024, \apj, 975, 1, 41. doi:10.3847/1538-4357/ad72ef

\bibitem[Zhou et al.(2025)]{Zhou2025} Zhou, S., Sun, M., Feng, H.-C., et al.\ 2025, \apj, 986, 2, 137. doi:10.3847/1538-4357/add468

\bibitem[Zu et al.(2011)]{Zu2011} Zu, Y., Kochanek, C.~S., \& Peterson, B.~M.\ 2011, \apj, 735, 2, 80. doi:10.1088/0004-637X/735/2/80

\end{thebibliography}
\end{document}